\documentclass[conference]{IEEEtran}
\IEEEoverridecommandlockouts
\usepackage{cite}
\usepackage{amsmath,amssymb,amsfonts}
\usepackage{physics}
\usepackage{algorithm}
\usepackage{algpseudocode}

\usepackage{graphicx}
\usepackage{textcomp}
\usepackage{xcolor}
\usepackage{booktabs} 
\usepackage{subcaption}
\usepackage{url}

\usepackage{newfloat}
\usepackage{listings}
\definecolor{codegreen}{rgb}{0,0.6,0}
\definecolor{codegray}{rgb}{0.5,0.5,0.5}
\definecolor{codepurple}{rgb}{0.58,0,0.82}
\definecolor{backcolour}{rgb}{0.95,0.95,0.92}
\lstdefinestyle{mystyle}{
    backgroundcolor=\color{backcolour},   
    commentstyle=\color{codegreen},
    keywordstyle=\color{magenta},
    numberstyle=\tiny\color{codegray},
    stringstyle=\color{codepurple},
    basicstyle=\ttfamily\footnotesize,
    breakatwhitespace=false,         
    breaklines=true,                 
    captionpos=b,                    
    keepspaces=true,                 
    numbers=left,                    
    numbersep=5pt,                  
    showspaces=false,                
    showstringspaces=false,
    showtabs=false,                  
    tabsize=2
}

\def\BibTeX{{\rm B\kern-.05em{\sc i\kern-.025em b}\kern-.08em
    T\kern-.1667em\lower.7ex\hbox{E}\kern-.125emX}}
\begin{document}

\title{An Efficient Split Fine-tuning Framework for Edge and Cloud Collaborative Learning
}

\author{\IEEEauthorblockN{Shaohuai Shi\IEEEauthorrefmark{2}, Qing Yang\IEEEauthorrefmark{3}, Yang Xiang\IEEEauthorrefmark{3}\IEEEauthorrefmark{1}\thanks{*Corresponding author.}, Shuhan Qi\IEEEauthorrefmark{2}, Xuan Wang\IEEEauthorrefmark{2}\\}
\IEEEauthorblockA{\IEEEauthorrefmark{2}Harbin Institute of Technology, Shenzhen, China
	\\\IEEEauthorrefmark{3}Peng Cheng Laboratory, Shenzhen, China\\
	shaohuais@hit.edu.cn, yq5427@163.com, xiangy@pcl.ac.cn, \{shuhanqi, xuanwang\}@cs.hitsz.edu.cn
	}
}

\maketitle

\begin{abstract}
To enable the pre-trained models to be fine-tuned with local data on edge devices without sharing data with the cloud, we design an efficient split fine-tuning (SFT) framework for edge and cloud collaborative learning. We propose three novel techniques in this framework. First, we propose a matrix decomposition-based method to compress the intermediate output of a neural network to reduce the communication volume between the edge device and the cloud server. Second, we eliminate particular links in the model without affecting the convergence performance in fine-tuning. Third, we implement our system atop PyTorch to allow users to easily extend their existing training scripts to enjoy the efficient edge and cloud collaborative learning. Experiments results on 9 NLP datasets show that our framework can reduce the communication traffic by 96 times with little impact on the model accuracy.
\end{abstract}


\section{Introduction}
In recent years, pre-trained language models (PLMs) (or called foundation models~\cite{bommasani2021opportunities})~\cite{kenton2019bert,floridi2020gpt,narayanan2021efficient,chowdhery2022palm} have achieved significant breakthroughs in many downstream natural language processing (NLP) applications like text generation~\cite{chen2020distilling}, language translation~\cite{adelani-etal-2022-thousand}, etc. Once PLMs are well trained with pre-training, they can be used in many scenarios with fine-tuning, where the model is fine-tuned on task-specific datasets with only several epochs going through the datasets. Compared to training from scratch, fine-tuning on PLMs normally takes much faster time (i.e., several epochs on task-specific datasets) and higher accuracy, so it becomes a common practice in many NLP tasks. For example, top-ranked models on GLUE~\cite{wang2018glue} and SQuAD~\cite{rajpurkar2016squad} benchmarks are fine-tuned from PLMs.

On the other hand, with the exponential growth of edge devices (e.g., mobile and Internet of Things, IoT), lots of generated data are privacy sensitive, which could not be shared for training models. The fine-tuning technique is very suitable for local training on edge devices as they can keep their data private to learn a model for local usage. However, due to the low computational resources and the memory limitation of edge devices, keeping all training processes on devices may not be possible or it takes very long training time. 
To alleviate this problem, split learning (SL)~\cite{vepakomma2018split} has become a promising distributed learning paradigm to enable resource-constraint edge devices to train deep neural networks (DNNs) with the help of powerful cloud servers without exposing their data to the server~\cite{oh2022locfedmix}. Specifically, SL splits the DNN into two parts (one part is stored on the edge and the other part on the cloud) at a particular layer. Meanwhile, modern DNN training (or fine-tuning) mainly uses stochastic gradient descent (SGD) and its variants (e.g., Adam) with backpropagation~\cite{hecht1992theory} to update model parameters iteratively. At each iteration, the training algorithm loads a mini-batch of local data to do the feed-forward computations, and then do backpropagation computations to calculate the gradients to update the model parameters. As the model is split into two parts in SL, the client should send the activation outputs to the server in the feed-forward pass, and the server sends the gradient w.r.t. the activation to the client in the backpropagation pass for updating model parameters locally~\cite{vepakomma2018split,palanisamy2021spliteasy}.

However, due to the bandwidth between the edge devices and the server (e.g., 1-1000Mb/s) is typically much smaller than the bandwidth between two servers (e.g., 1-200Gb/s) in a data center, the data exchange of communicating activation outputs and their gradients between the client and the server is very slow. For example, fine-tuning a popular BERT$_{BASE}$ model (around 110 million parameters)~\cite{kenton2019bert} on an Nvidia V100 GPU takes 120ms per iteration, while the communication volume per iteration is 340MB which requires 2300ms for communication in SL under a 1000Mb/s connection. It means that the introduced communication cost hinders the advantage of SL in making use of powerful servers for training. While there exist some studies~\cite{gao2020compressing,cai2020tinytl,liu2021enabling} trying to reduce training or fine-tuning costs on edge devices, they fail to address the communication problem in SL.

To this end, in this work, we propose an efficient split fine-tuning framework, SFT. Specifically, we first identify the low-rank property of weight parameters and their gradients in fine-tuning BERT models (\S\ref{subsec:convergence}). Then, we propose a novel compression approach (\S\ref{sec:system}) to reduce the communication cost of exchanging data between the edge device and the cloud server by decomposing a single feed-forward layer into three much smaller feed-forward layers based on matrix decomposition while \textit{requiring no extra computation overheads}. We implement our framework\footnote{Our code is available in \url{https://openi.pcl.ac.cn/Encore/splitfinetuning}.} atop PyTorch to allow users to easily utilize our SFT with pre-trained models with little impact on model accuracy (\S\ref{subsec:implementation}). To show the effectiveness of our SFT, we conduct extensive experiments on GLUE~\cite{wang2018glue} and SQuAD~\cite{rajpurkar2016squad} datasets with the pre-trained BERT~\cite{kenton2019bert} model. Experimental results show that SFT can reduce 96$\times$ communication volume than SL with little impact on the model accuracy. 

The rest of the paper is organized as follows. We first introduce some background and related work in Section~\ref{sec:background}. Then we present our proposed system in Section~\ref{sec:system}. After that, we demonstrate the experimental studies in Section~\ref{sec:evaluation}. We finally conclude the paper in Section~\ref{sec:conclusion}.

\begin{figure*}[!t]
	\centering
	\begin{subfigure}{0.46\textwidth}
		\includegraphics[width=\linewidth]{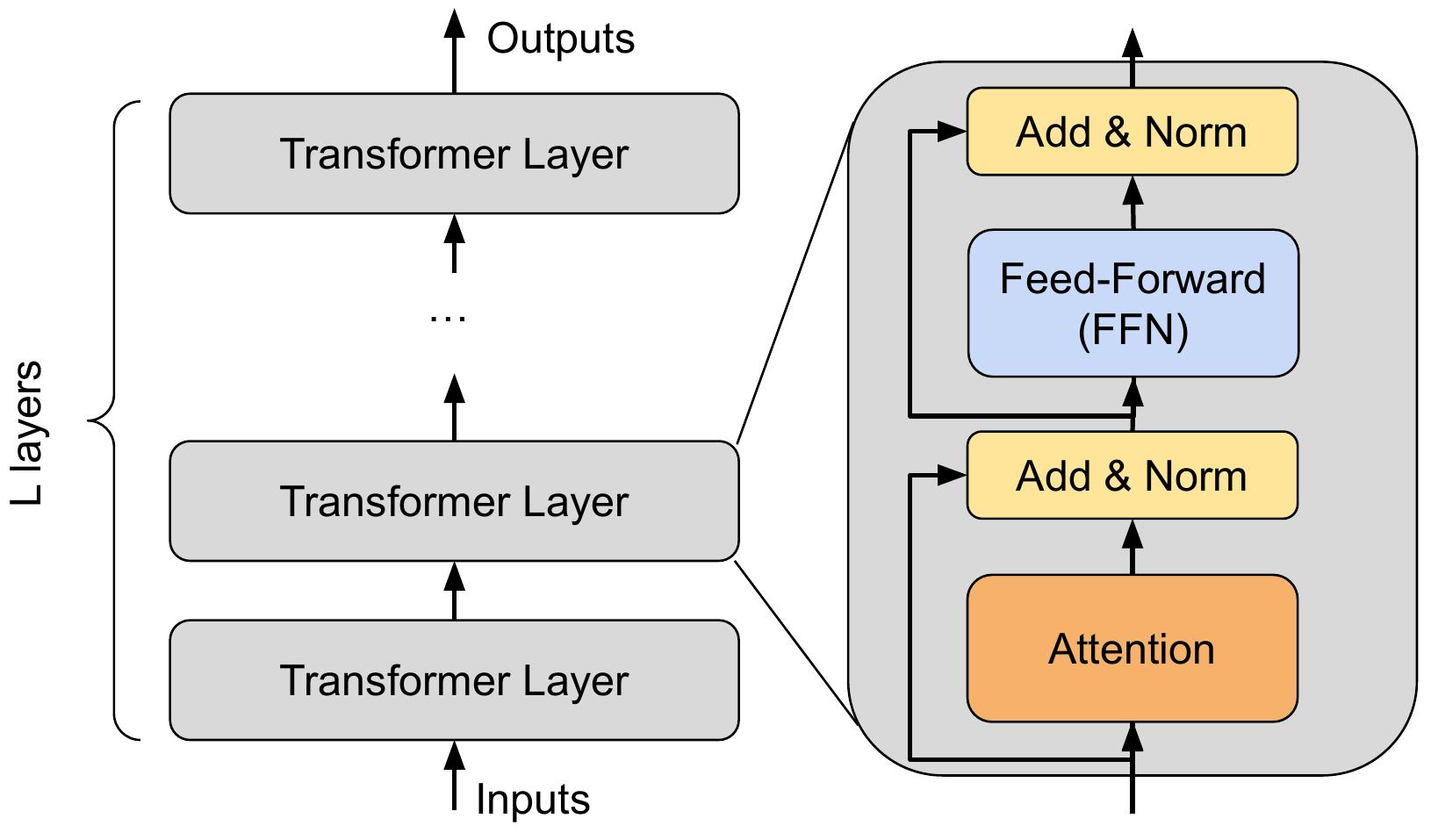}
		\caption{Transformer based pre-train model.}
	\end{subfigure}
	\hspace{4pt}
	\begin{subfigure}{0.22\textwidth}
		\includegraphics[width=\linewidth]{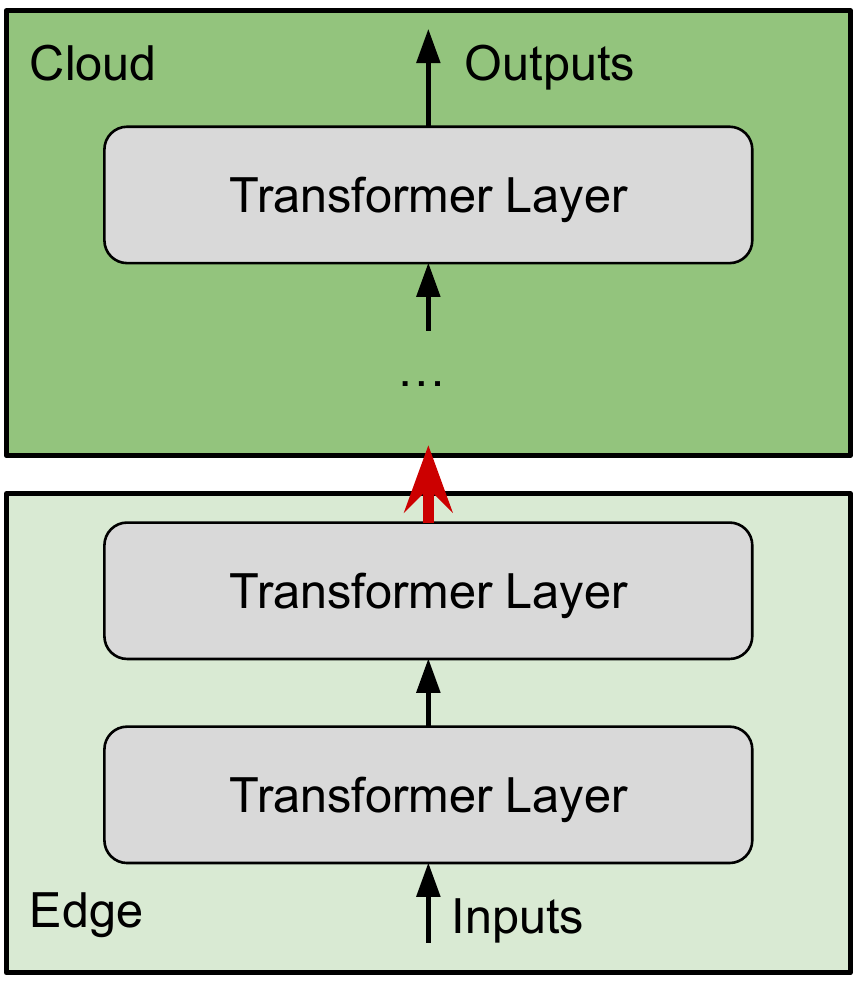}
		\caption{Split learning.}
	\end{subfigure}
	\hspace{4pt}
	\begin{subfigure}{0.22\textwidth}
		\includegraphics[width=\linewidth]{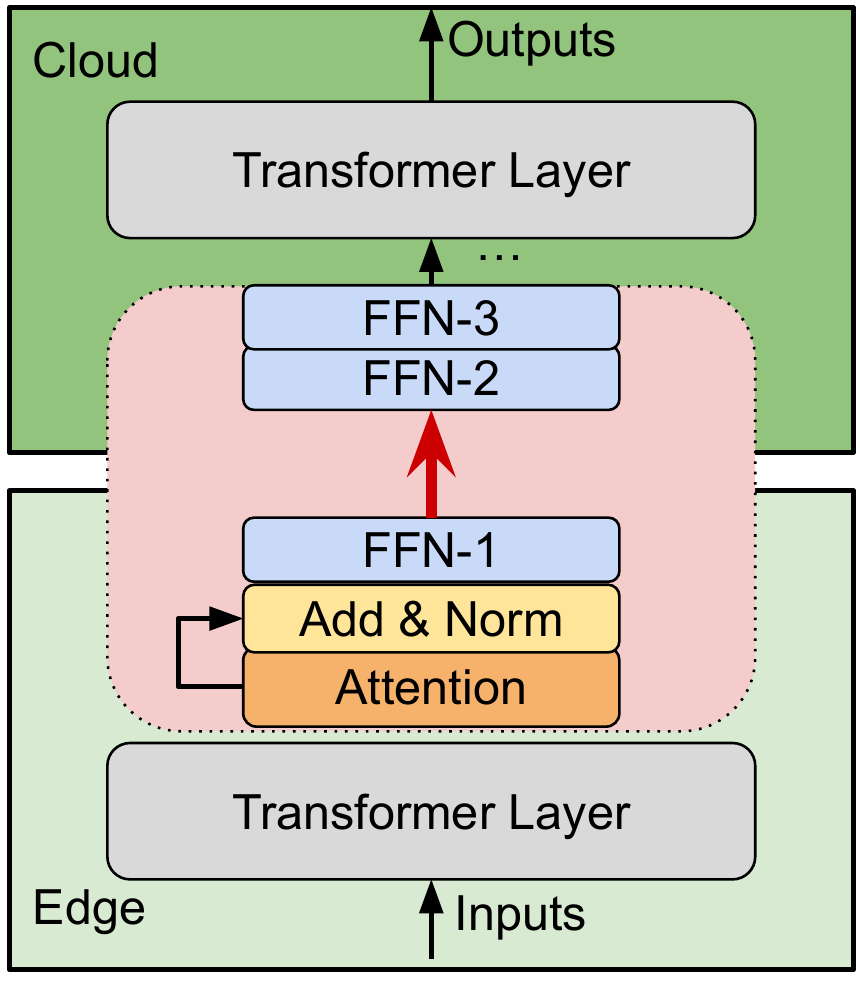}
		\caption{Split fine-tuning.}
	\end{subfigure}
	\caption{(a) A typical transformer architecture which consists of multiple transformer layers (say L layers), and each transformer layer (also called a block) has a feed-forward (FFN) layer. (b) Split learning architecture: the full model is split into two parts, each of which is put on the edge and the cloud respectively. (c) Our split fine-tuning framework (SFT): the split layer (FFN) is decomposed into three smaller FFNs, i.e., FFN-1 is located on the edge; FFN-2 and FFN-3 are located on the cloud.}
	\label{fig:full-figure}
\end{figure*}

\section{Background and related Work}\label{sec:background}
\subsection{The feed-forward layer (FFN) in Transformer}
Existing popular PLMs in NLP are mainly built on the Transformer~\cite{ashish2017attention} architecture. As shown in Fig.~\ref{fig:full-figure}(a), a PLM consists of a series of transformer layers (or blocks), each of which contains an attention layer and a feed-forward layer (FFN) with residual connections. Since each layer requires a residual connection (Add \& Norm), the input size and the output size of the layer should be identical, which means the shapes of all input and output tensors within the transformer layer or between transformer layers are identical. This is an important feature causing extensive communication traffics in SL for fine-tuning such models. See below for more details.

\subsection{Split learning}
Split learning (SL)~\cite{vepakomma2018split,poirot2019split} has provided an emerging solution for edge and cloud collaborative learning without exposing the private data on the edge while enjoying the powerful computational resources from the cloud servers. As shown in Fig.~\ref{fig:full-figure}(b), a DNN model is split into two parts at a particular layer. One part (say $net_1$) is stored on the edge and the other part (say $net_2$) is stored on the cloud. The number of layers on each part may be different, and more layers typically take higher computational costs. Thus, in SL, the data located on the edge is loaded for feed-forward computation on $net_1$, and its activation output is transferred (\textit{upload}) to the cloud. We assume the activation output is a tensor $\vb*{a}$ with shape $(B, M, N)$, where $B$ is the mini-batch size and $M$ and $N$ are two dimensions of a matrix of the hidden feature representation. Note that $M$ and $N$ may be the explicit dimensions of the hidden feature as different types of layers (e.g., FFN, attention, etc.) have different structures, but they all can be organized as a matrix. For the transformer-based architecture, as we analyzed in the above subsection, the shape is the same no matter which layer is chosen as the split layer in SL. During the backpropagation pass, the gradient w.r.t. $\vb*{a}$, which is denoted as $\vb*{\delta}$ and has the same shape with $\vb*{a}$, is calculated on the cloud side and transferred to the edge for calculating the gradients w.r.t. the model parameters for updating the model. 

It is seen that the computations of training have been partially ($net_2$) offloaded to the server to utilize the powerful server to reduce the training time for edge devices. However, the exchange data (i.e., $\vb*{a}$ and $\vb*{\delta}$) has a large number of elements, which requires significant communication time at each training iteration and becomes the performance bottleneck. In this work, we will present how to reduce the communicating volume of $\vb*{a}$ and $\vb*{\delta}$ without sacrificing model accuracy.

\subsection{Activation or model compression}
Some works in reducing $\vb*{a}$ and $\vb*{\delta}$ are activation compression techniques~\cite{georgiadis2019accelerating,evans2021ac,xia2022structured,liu2021enabling,xiang2022nebula}. These studies try to compress the activation outputs for each layer to save computational costs and memory for storing the temporary data in activation, which can be classified as the quantization method and the pruning method with tensor decomposition. Particularly, Evans et al.,~\cite{evans2021ac} propose the AC-GC lossy compression algorithm to dynamically compress the activation via quantization and an optimization objective of maximizing the compression ratio while minimizing the accuracy loss. While the quantization technique can maximally reduce the storage of $\vb*{a}$ (with 1-bit) by 32 times compared to the 32-bit counterpart, too aggressive compression easily introduces an accuracy loss in model training~\cite{evans2021ac} in practice. The tensor decomposition approaches~\cite{gao2020compressing,liu2021enabling,xia2022structured} are to decompose the parameter tensors to small-size approximate tensors, thus reducing the computational and memory costs. Particularly, MPO~\cite{gao2020compressing, liu2021enabling} decomposes an original parameter matrix (say $\vb*{w}$) into $n$ ($n\geq 2$) small matrices (i.e., $\vb*{w}\rightarrow [\vb*{w}_1, \vb*{w}_2, ..., \vb*{w}_n]$). Even though MPO achieves high compression ratios with slight accuracy loss, it requires particular tricks like dimension squeezing and takes extra computational costs for finding the layer with the least reconstruction error, which makes the optimization procedure difficult to be applied in edge and cloud collaborative learning. 

\section{SFT: The Split Fine-tuning Framework}\label{sec:system}
\subsection{Overview}
The architecture overview of our proposed SFT is shown in Fig.~\ref{fig:full-figure}(c). The key idea is two folds: 1) decomposing the FFN layer into three smaller FFN layers (FFN-1, FFN-2, and FFN-3) after loading the pre-trained parameters, and 2) the residual connection in the original FFN layer is eliminated, but the model convergence is guaranteed. The communication volume between the edge and cloud in the split layer of FFN-1 becomes much smaller than the original FFN output, thus significantly reducing the communication time in collaborative learning. The details are provided as follows.

\subsection{System architecture}
As the high communication cost is caused by the large dimension of $\vb*{a}$ and $\vb*{\delta}$ at each iteration, we propose to decompose $\vb*{w}$ to multiple small tensors, so that the generated activation output is with small dimensions. Specifically, given a transformer-based model, we first load the pre-trained parameters to the original model architecture. In SFT, we only need to compress the layer that needs to be communicated between the edge and the cloud. According to our observation that decomposition does not affect the model accuracy (\S\ref{subsec:convergence}), the weights in FFN layers in fine-tuning PLMs are mostly low-rank matrices. We choose the FFN layer as the split layer in SL. Formally, for an FFN layer at layer $l$ with weight matrix $\vb*{w}_l$ and the input $\vb*{a}_{l-1}$ (the output of its previous layer), its output can be represented as 
\begin{equation}\label{equ:ffn-computation}
    \vb*{a}_l = \vb*{a}_{l-1}\vb*{w}_l.
\end{equation}
Assuming that SL splits the DNN into two parts at layer $l$, $\vb*{a}_l\in \mathbb{R}^{M\times N}$ should be communicated from the edge to the cloud in the forward pass, and its gradient $\vb*{\delta}_l\in \mathbb{R}^{M\times N}$ should be communicated from the cloud to the edge in the backward pass. Note that in fine-tuning tasks, $\vb*{w}_l$ has been initialized with pre-trained parameters. Our goal is to compress $\vb*{a}_l$ and $\vb*{\delta}_l$ such that the communication volume is small enough to eliminate the communication bottleneck in SL.

In SFT, we decompose the split layer (i.e., layer $l$) whose weights are denoted by $\vb*{w}\in \mathbb{R}^{N\times H}$ to three matrices via singular value decomposition (SVD), i.e.,
\begin{equation}
    \vb*{w}=\vb*{u}\Sigma \vb*{v},
\end{equation}
where $\vb*{u}\in \mathbb{R}^{N\times R}$, $\Sigma \in \mathbb{R}^{R\times R}$ is a diagonal matrix whose diagonal entries are singular values of $\vb*{w}$, and $\vb*{v} \in \mathbb{R}^{R\times H}$. $R\leq \min\{N, H\}$ is the rank of 
$\vb*{w}$. With the decomposed matrices, Eq. (\ref{equ:ffn-computation}) becomes
\begin{equation}\label{equ:decomposition}
    \vb*{a}_l = \vb*{a}_{l-1}\vb*{u}\Sigma \vb*{v}=(\vb*{a}_{l-1}\vb*{u})(\Sigma\vb*{v}).
\end{equation}
Let $\hat{\vb*{a}}_l=\vb*{a}_{l-1}\vb*{u}$ and $\hat{\vb*{a}}_l\in \mathbb{R}^{M\times R}$. Note that $\hat{\vb*{a}}_l=\vb*{a}_{l-1}\vb*{u}$ is equivalent to constructing an FFN layer whose weight is $\vb*{u}$. Similarly, $\vb*{a}_l=\hat{\vb*{a}}_l(\Sigma\vb*{v})$ is equivalent to constructing two FFN layers whose weights are $\Sigma$ and $\vb*{v}$ respectively. Instead of splitting the DNN in the original architecture at layer $l$, SFT splits its decomposed form at $\vb*{u}$. It means that the activation output on the edge side is $\hat{\vb*{a}}_l\in \mathbb{R}^{M\times R}$, which should be communicated to the server side in the forward pass. The corresponding gradient is $\hat{\vb*{\delta}}_l\in \mathbb{R}^{M\times R}$ in the backward pass. Thus, the communication volume is reduced from $M\times N$ to $M\times R$, i.e., the communication time is shortened by $N/R$ times theoretically. 

Note that SFT decomposes a single FFN layer to three smaller FFN layers after loading the pre-trained parameters but before fine-tuning. The three constructed FFN layers are initialized with the SVD decomposition and they are tuned every iteration in fine-tuning. Due to the low-rank feature of the weight matrix of the FFN layer during fine-tuning, we can choose an extremely small value of $R$ without affecting the convergence performance in fine-tuning. In our experiments, setting $R=1$ or $R=8$ can almost preserve the model accuracy (\S\ref{sec:evaluation}). In summary, SFL can reduce the communication traffic by $N/R$ times over SL.

\subsection{Algorithm}
\begin{algorithm}[!ht]
	\caption{SFT: Split fine-tuning on an edge and a cloud}\label{algo:sft}
 	\small
		\textbf{Input: }$net, l, I$ 
	\begin{algorithmic}[1]
	    \State Split $net$ to $net_1$ and $net_2$ at layer $l$;
	    \State Load pre-trained parameters \Comment{$net_1$ on edge and $net_2$ on cloud};
	    \State Reconstruct layer $l$ with three FFN layers with Eq. (\ref{equ:decomposition}); \Comment{Keep the first layer on edge and keep the last two layers on cloud};
	    \For{$i = 1\rightarrow I$}
	        \State Load $B$ training samples: $(\vb*{x}_i, \vb*{y}_i)$; \Comment{On edge}
	        \State Feed-forward with $net_1$: $\hat{\vb*{a}}_l=net_1(x_i)$; \Comment{On edge}
	        \State Edge sends $\hat{\vb*{a}}_l$ and $y_i$ to Cloud;
	        \State Feed-forward with $net_2$: $\hat{\vb*{y}}_i=net_2(\hat{\vb*{a}}_l)$; \Comment{On cloud}
	        \State Calculate $loss$ with $\hat{\vb*{y}}_i$ and $\vb*{y}_i$; \Comment{On cloud}
	        \State Back-propagation with loss: $\hat{\vb*{\delta}}_l=loss.backward()$; \Comment{On cloud}
	        \State Cloud sends $\hat{\vb*{\delta}}_l$ to Edge;
	        \State Back-propagation with $\hat{\vb*{\delta}}_l$; \Comment{On edge}
	        \State Update $net_1$; \Comment{On edge}
	        \State Update $net_2$; \Comment{On cloud}
	    \EndFor
    \end{algorithmic}
\end{algorithm}
The SFT algorithm is shown in Algorithm~\ref{algo:sft}, where some comments illustrate that some code is only executed on the edge and some are only on the cloud. The inputs are the neural network architecture $net$, the split layer $l$, and the number of iterations $I$ for fine-tuning. In Algorithm~\ref{algo:sft}, lines 1-3 are splitting the model, loading pre-trained models, and reconstructing the split layer based on SVD. The for loop in lines 4-14 is the training procedure in both the edge and cloud sides. For each iteration, the edge first loads the training data (line 5) and then performs the feed-forward on the first part of the network (line 6), whose results ($\hat{\vb*{a}}$) are sent to the cloud (line 7). After the cloud receives the activation from the client, it performs feed-forward on the second part of the network (line 8), followed by the backward computation with loss (lines 9-10). The gradient ($\hat{\vb*{\delta}}$) w.r.t. the activation is sent from the cloud to the edge (line 11), followed by back-propagation on the edge side (line 12). After that, the models are updated simultaneously on the edge and the cloud (lines 13-14). Our system enables users with existing training scripts to be able to explore the cloud to fine-tune their models without exposing the data to the server.

\subsection{Performance analysis}
SFL has two main goals: 1) making the low-memory edge device possible for fine-tuning when the edge cannot store the whole model, and 2) enabling edge devices to explore the cloud servers to fine-tune a model more efficiently. The first goal is obviously our advantage using SFT when low-memory devices cannot store the whole model. For the second goal, however, we should consider whether SFT can reduce the overall fine-tuning time. Assume that the original iteration time on the edge to fine-tune a model $net$ is denoted by $t_{naive}=t_{edge}(net)$. When we use SFT, the fine-tuning time per iteration is 
\begin{equation}
    t_{sft}=t_{edge}(net_1)+t_{cloud}(net_2)+t_{comm},
\end{equation}
where $t_{cloud}(net_2)$ is the computation time on the cloud and $t_{comm}$ is the communication time between the edge and the cloud. Thus, SFT can be used if $t_{sft}<t_{naive}$. As the cloud server typically has much higher computational power than the edge device, $t_{edge}(net_1)+t_{cloud}(net_2)$ should become much smaller than $t_{naive}$. However, $t_{comm}$ should not be too large so that one can enjoy the efficiency of SFT. 

$t_{edge}(net_1)$ and $t_{cloud}(net_2)$ depend on the split layer. A lower split layer indicates that more computational workloads are uploaded to the server for calculation, which would make $t_{edge}(net_1)$ smaller and $t_{cloud}(net_2)$ larger, and vice visa. As the cloud is much more powerful than the edge, we expect the split layer to be as low as possible, but the lower layer may sacrifice the model accuracy (\S\ref{subsec:convergence}). Therefore, we have a trade-off between accuracy and efficiency in SFT. $t_{comm}$ depends on the rank we used for decomposition. A larger rank makes $t_{comm}$ higher. Therefore, the two parameters (i.e., split layer and rank) should be well-tuned for achieving better end-to-end training performance in SFT. In this work, we do not develop a strategy for tuning them, but we present some observations from the experiments for helping tune them.

\subsection{Implementation}\label{subsec:implementation}
To enable users to easily use our SFT framework, we implement a distributed optimizer named ``SFTOptimizer'' atop the PyTorch optimizer to integrate the layer decomposition of the model and the communication between the edge and the cloud. To use SFT, users only need two more lines of code to extend their original training scripts. An example is shown in Listing~\ref{samplecode}, where lines 2-3 are inserted into the original training code to use SFT. 
\begin{lstlisting}[label=samplecode,caption=An example of SFT usage,language=Python]
optim = torch.optim.Adam(model.parameters(), ...)
role = 'edge' if edge else 'cloud' # +++
optim = SFLOptimizer(optim, role=role, ...) # +++ 
for i in range(epochs):
    for data, target in train_loader:
        optim.zero_grad()
        output = model(data)
        loss = criterion(output, target)
        loss.backward()
        optim.step()
        ...
\end{lstlisting}

\vspace{5pt}
\section{Evaluation}\label{sec:evaluation}
\subsection{Experimental settings}
\textbf{Testbed. }We use two Nvidia Tesla V100 GPUs in a single node as an emulation environment. One GPU is used to simulate an edge device, and one GPU is used to work as a cloud server.

\textbf{DNN and datasets. }We use the popular language model, BERT$_{BASE}$~\cite{kenton2019bert}, which has 12 layers and 110M parameters, as our experimental neural network. The dimension of the split layer is $(M, N)=(3072, 768)$. Its corresponding pre-trained model is downloaded from the well-trained parameters at HuggingFace\footnote{\url{https://huggingface.co/bert-base-uncased}}. We fine-tune the model on 9 datasets from GLUE~\cite{wang2018glue} and SQuAD~\cite{rajpurkar2016squad} for different downstream tasks including named entity recognition, textual entailment, co-reference resolution, etc.

\subsection{Convergence performance}\label{subsec:convergence}
\begin{figure}[!t]
	\centering
	\begin{subfigure}{0.24\textwidth}
		\includegraphics[width=\linewidth]{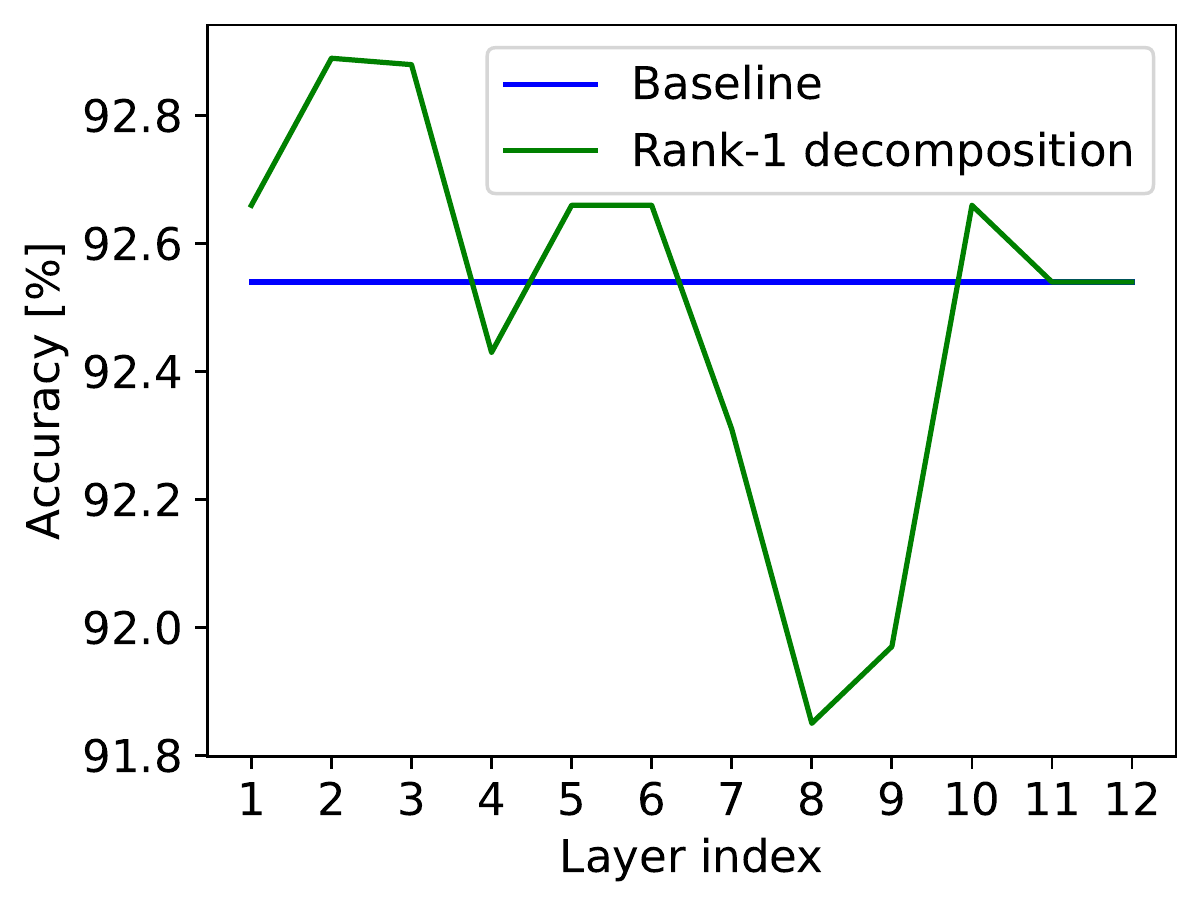}
		\caption{SST-2.}
	\end{subfigure}
	\begin{subfigure}{0.24\textwidth}
		\includegraphics[width=\linewidth]{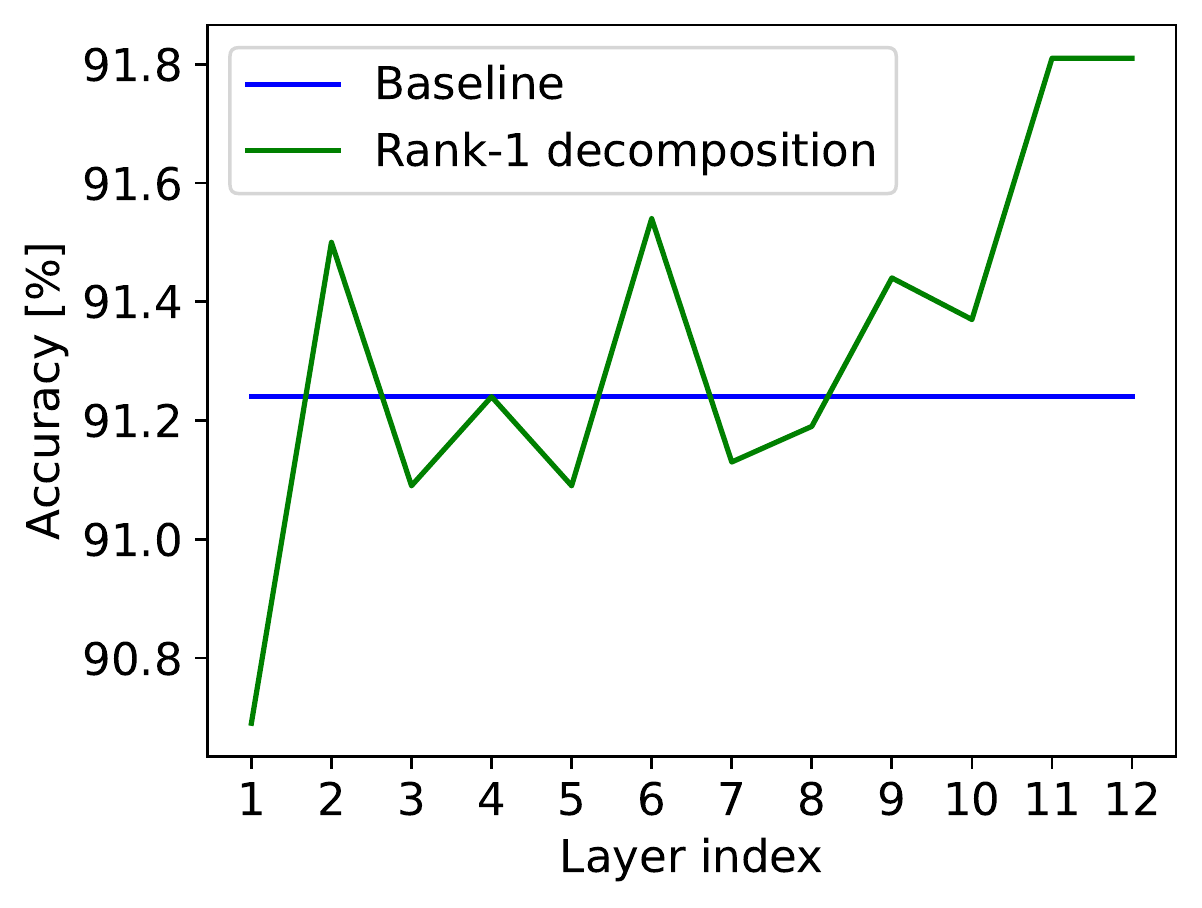}
		\caption{QNLI.}
	\end{subfigure}
	\caption{Model accuracy on validation tests w/ rank-1 decomposition in SFT while preserving the residual connection. The baseline is the result run with the original fine-tuning algorithm.}
	\label{fig:convergence-withres}
\end{figure}

\begin{figure}[!t]
	\centering
	\begin{subfigure}{0.24\textwidth}
		\includegraphics[width=\linewidth]{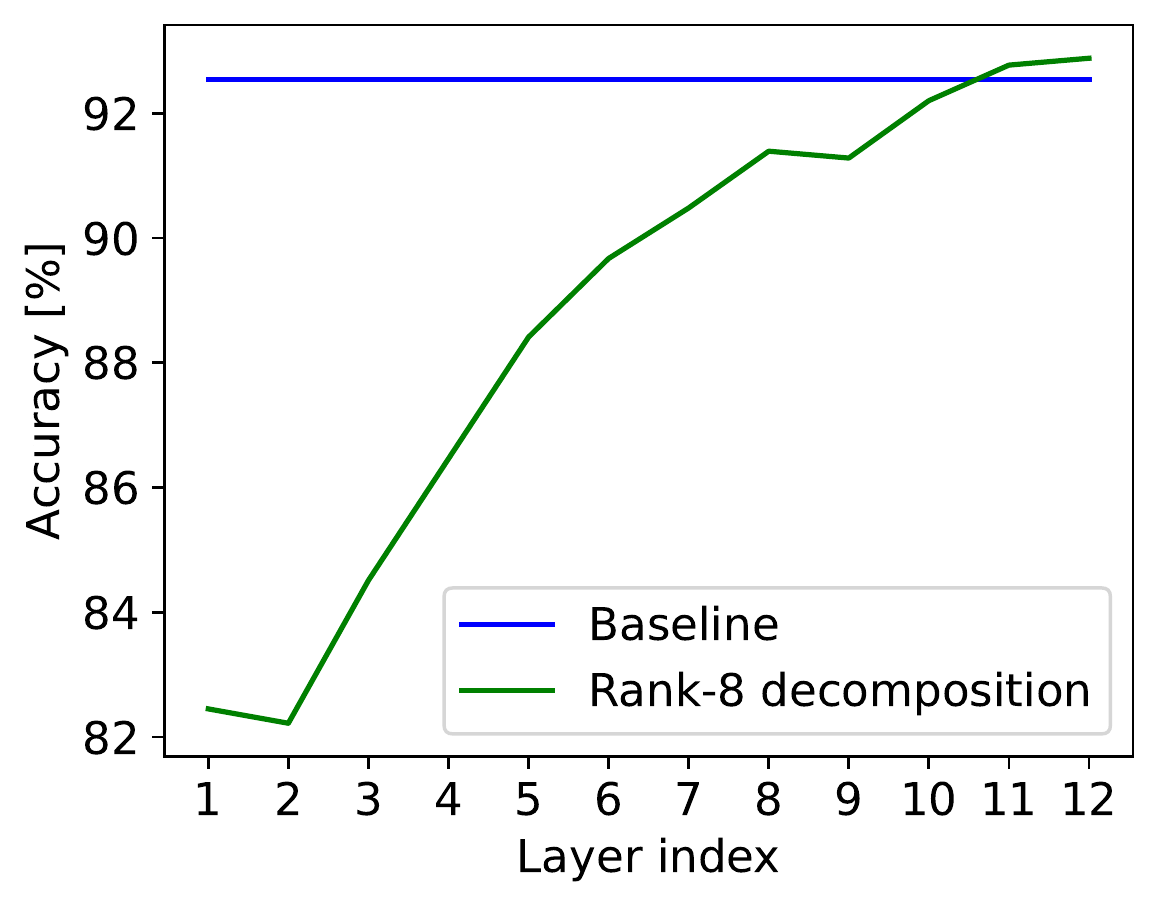}
		\caption{SST-2.}
	\end{subfigure}
	\begin{subfigure}{0.24\textwidth}
		\includegraphics[width=\linewidth]{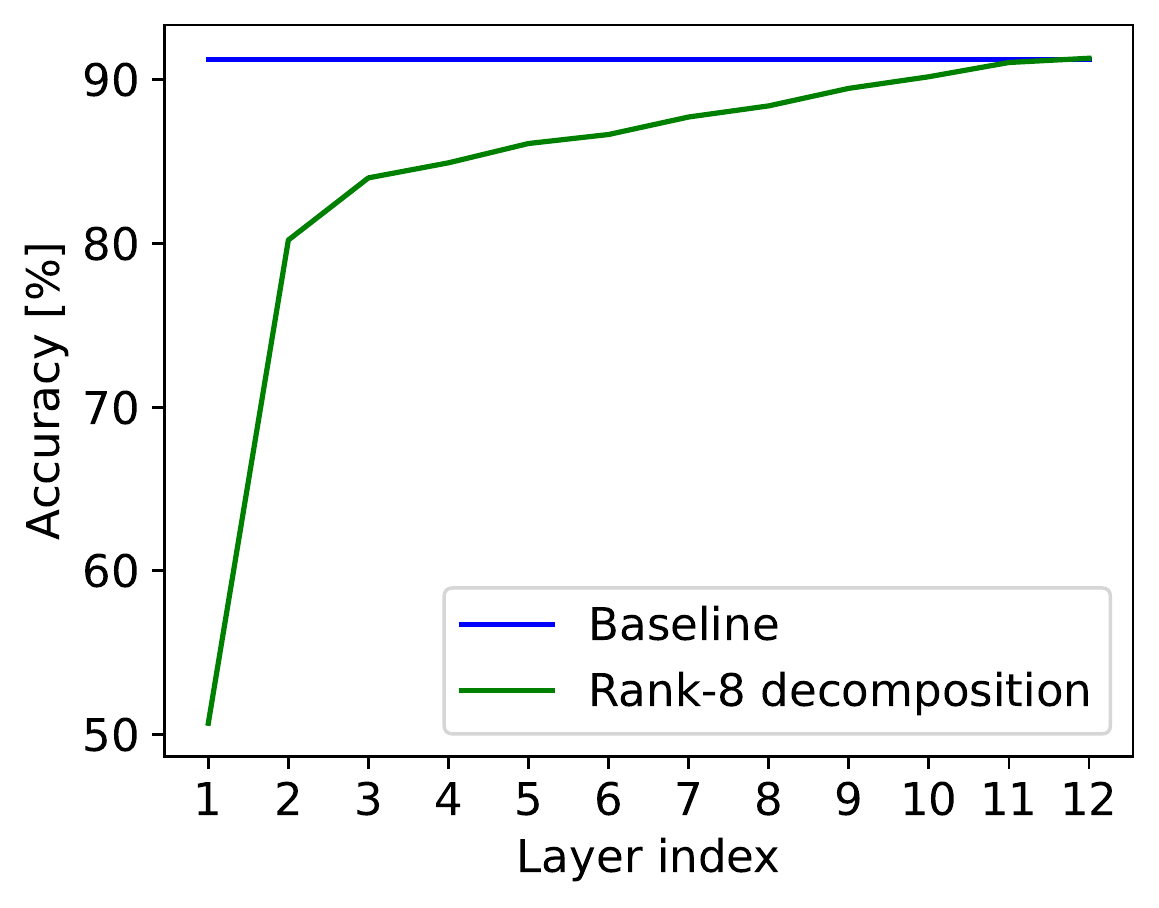}
		\caption{QNLI.}
	\end{subfigure}
	\caption{Model accuracy of on validation tests w/ rank-8 decomposition in SFT while eliminating the residual connection.}
	\label{fig:convergence-withoutres}
\end{figure}
\textbf{SFT w/ the residual connection. }We first demonstrate the convergence results of the SVD decomposition for replacing the large FFN with three small FFNs for fine-tuning, which shows the low-rank feature of FFN weights. We use SST-2 and QNLI datasets to study the convergence property of our SFT as shown in Fig.~\ref{fig:convergence-withres}. Due to the page limit, we do not show the results of other datasets as they have similar patterns. The results show that the weight matrix in the split layer can be decomposed as a rank-1 matrix without sacrificing the model accuracy. In some scenarios, rank-1 decomposition is better than the baseline. For example, decomposing at layers 2 or 3 (11 or 12)\footnote{The smaller layer index means the lower layer} is much better than the baseline in SST-2 (QNLI). How to determine which layer should be split to achieve the highest efficiency in a given environment is also of importance to explore, and we will leave it as our future work.

\textbf{SFT w/o the residual connection. }Since we also need to eliminate the residual connection in the original transformer layer in SFT, we conduct experiments to verify the convergence performance by eliminating the residual connection and decomposing the FFN with SVD. The results are shown in Fig.~\ref{fig:convergence-withoutres}, which shows that the model accuracy tends to decrease when the split layer is chosen at lower layers. However, it can still preserve the model accuracy when splitting at higher layers. Thus, it is possible to use SFT to enjoy the powerful cloud server to accelerate the fine-tuning tasks for edge devices without sharing the data.

\begin{table*}[!ht]
  \centering
  \caption{Fine-tuning accuracy (the higher the better) in validation sets. For each dataset (the numbers in brackets indicate the sizes of the datasets), we run three independent experiments and calculate their mean.}
  \label{table:convergence-perf}
    \begin{tabular}{cccccccccc}
    \toprule
    Algorithm & SST-2 & QNLI & MNLI & QQP & CoLA & RTE & STS-B & MRPC &SQuAD\cr
              & (67k) & (105k) & (364k) & (91.2k) & (8.5k) & (2.5k) & (7k) &(3.7k) & (88k) \cr\hline
    Baseline & 92.54 & 91.24 & 84.56 & 90.73 & 55.3  & 66.06 & 88.38 & 85.33  & 88.25 \cr
    SFT(l=11,R=8) & 92.43 & 90.98 & 83.98 & 90.93 & 57.13 & 64.25 & 86.46 & 84.81 & 88.33 \cr
    SFT(l=11,R=16) & 92.31 & 91.22 & 84.33 & 90.75 & 57.35 & 62.09 & 86.78 & 83.47 & 88.75 \cr
    SFT(l=11,R=32) & 92.77 & 91.04 & 84.27 & 90.99 & 57.87 & 62.81 & 87.46 & 84.23 & 88.56 \cr
    \bottomrule
    \end{tabular}
\end{table*}

\textbf{Model accuracy in all datasets. }Based on the convergence results in Fig.~\ref{fig:convergence-withoutres}, we use a rank of 8, 16, and 32 for decomposition at layer 11 in SFT (denoted as SFT(l=11,R=8), SFT(l=11,R=12), and SFT(l=11,R=32) respectively). Therefore, the communication cost can be reduced by 768/8=96 times in the chosen BERT model if R=8. The model accuracy of validation sets in all chosen 9 datasets is shown in Table~\ref{table:convergence-perf}. The results show that different ranks have little differences and a higher rank does not guarantee a higher accuracy. For example, SFT(l=11,R=8) achieves higher accuracy than SFT(l=11,R=16) on QQP and RTE datasets. In most cases, SFT(l=11,R=8) preserves the model accuracy compared to the baseline. In the particular case of RTE, SFT(l=11,R=8) achieves much lower accuracy than the baseline due to the extremely small size (only around 2,500 training samples) of the dataset. The results show that SFT (using a rank of 8) makes the fine-tuning tasks possible to train in a collaborative learning environment so that the edge does not need to expose the data to the server.


\subsection{Estimated iteration performance}
To show the end-to-end performance of our SFT compared with SL on the BERT$_{BASE}$ model, we benchmark the time performance on a V100 GPU using SST-2 (other datasets have similar patterns). Let $t_{bert}(gpu, nlayers)$ denote the wall-clock time per iteration training the BERT$_{BASE}$ model on a particular $gpu$ using $nlayers$ layers of the model. The full 12 layer BERT$_{BASE}$ on SST-2 takes 
\begin{equation}
t_{bert}(\text{V100}, 12)=124\text{ms}
\end{equation}
with a mini-batch size of 32 and a sequence length of 66. Since the 12 layers of BERT$_{BASE}$ have an identical architecture, each layer has the same computational workloads and thus has the same computation time. We can estimate the iteration time of one layer as 
\begin{equation}
t_{bert}(\text{V100}, 1)=124/12\text{ms}=10.3\text{ms}. 
\end{equation}
Assume that the edge side is a relatively new Nvidia edge device, XAVIER-NX, with 21 TOPs AI performance, and the cloud side is a V100 GPU with 130 TOPs AI performance, which means the cloud server is around 6 times faster than the edge device. Therefore, on a XAVIER-NX GPU, we have
\begin{equation}
t_{bert}(\text{XAVIER},12)=6t_{bert}(\text{V100}, 12)=6\times124=744\text{ms}
\end{equation}
and 
\begin{equation}
t_{bert}(\text{XAVIER},1)=6t_{bert}(\text{V100}, 1)=6\times10.3=60.3\text{ms}.
\end{equation}
The communication traffic with SL is $32\times 3076\times 768 \times 4=288$MB, while it is $32\times 3076\times 8 \times 4=36$MB with our SFT. Assume that the typical bandwidth between the edge to the cloud is 1000Mbps Ethernet\footnote{We also assume the bandwidth can be fully utilized, while it cannot in practice. The main purpose of the emulation is to demonstrate the potential benefits of our SFT in real-world environments, and the performance also varies with different configurations.}, the communication time of SL and SFT is 2,300ms and 24ms respectively. Thus, the iteration times with local training (on an edge device), with SL (on an edge and a cloud), and with SFT (on an edge and a cloud) are
\begin{equation}
    t_{naive}=t_{bert}(\text{XAVIER}, 12)=744\text{ms},
\end{equation}
\begin{align}
    t_{sl}&=t_{bert}(\text{XAVIER}, 10)+t_{bert}(\text{V100}, 2)+t_{comm} \notag\\
     & =(60.3\times 10+10.3\times 2+2300)\text{ms}=2923.6\text{ms},
\end{align}
and
\begin{align}
    t_{sft}&=t_{bert}(\text{XAVIER}, 10)+t_{bert}(\text{V100}, 2)+t_{comm}\\
    &=(60.3\times 10+10.3\times 2+24)\text{ms}=647.6\text{ms}
\end{align}
respectively. The results show that the introduced communication time in SL is very high, making it even slower than the local training. Our SFT, on the other hand, can achieve faster training time (14\% reduction in the end-to-end training time) by reducing the communication volume between the cloud server and the edge device.

\subsection{Discussion}
Our experimental results conclude two folds. First, when the edge devices cannot conduct fine-tuning tasks due to their memory constraint, it is possible to enjoy our SFT to fine-tune the model. Second, even if the edge clients can fine-tune the model locally, our SFT can help accelerate the training by exploring the more powerful cloud servers without requiring the data from the edge devices. However, there are still two important problems that should be further studied to make SFT more practical. First, could it be possible to split the layer in the lower layer of the model such that more computational workloads can be offloaded to the server? In our existing results, splitting the lower layer may introduce some accuracy loss. Thus, it is a trade-off between the model accuracy and training efficiency. Enjoying higher training speed may sacrifice some accuracy. Second, as keeping the residual connection can preserve the model accuracy (as shown in Fig.~\ref{fig:convergence-withres}) even using rank-1 decomposition, could it be possible to keep the residual connection without introducing significant communication costs between the edge and the cloud? As SFT decomposes the FFN to smaller FFNs which are distributed to the client and server, the residual connection should require the activation data to be transferred between the edge and the cloud, which makes the communication extremely heavy. However, eliminating the residual connection only allows higher layers to be split layers, which means only a small proportion of layers can be uploaded to the server if we want to preserve the model's accuracy.

\section{Conclusion}\label{sec:conclusion}
In this work, we proposed an efficient split learning framework for fine-tuning tasks, which is called split fine-tuning (SFT). Specifically, we first observed that model weights are normally low-rank in fine-tuning tasks based on our extensive experiments. Then based on the low-rank feature of fine-tuning, we introduced a novel layer decomposition method using SVD such that we can significantly reduce the communication volume between the edge and the cloud in collaborative learning. We implemented our prototype system atop PyTorch and enable end-users to easily conduct SFT with very little change to their existing training scripts. Extensive experimental results showed that our SFT reduces the communication traffic by 96 times compared to SL with little impact on the model accuracy.


\bibliographystyle{IEEEtran}
\bibliography{main}
\end{document}